\documentclass[a4paper]{jpconf}

\usepackage{amsmath,amssymb,amsfonts,amsthm}
\usepackage{euscript}
\usepackage{graphicx}

\begin{document}

\title{Topological geons with self-gravitating phantom scalar field}

\author{P V Kratovitch, I M Potashov, Ju V Tchemarina and A N Tsirulev}

\address{Faculty of Mathematics, Tver State University, 35 Sadovyi, Tver, Russia, 170002}

\ead{kratovitch.pv@tversu.ru, chemarina.yv@tversu.ru, potashov.im@tversu.ru, tsirulev.an@tversu.ru}

\begin{abstract}
A topological geon is the quotient manifold $M/Z_2$ where $M$ is a static spherically symmetric wormhole having the reflection symmetry with respect to its throat. We distinguish such asymptotically flat solutions of the Einstein equations according to the form of the time-time metric function by using the quadrature formulas of the so-called inverse problem for self-gravitating spherically symmetric scalar fields. We distinguish three types of geon spacetimes and illustrate them by simple examples. We also study possible observational effects associated with bounded geodesic motion near topological geons.
\end{abstract}

\section{Introduction}

A (topological) geon is an orientable asymptotically flat spacetime with topology $\mathbb{R}\!\!\;\times\!\!\; \mathbb{R}P^3\!\!\;\setminus\!\{p\}$~\cite{Sorkin1986}. The 'spatial' factor is the punctured real projective space and the omitted point is normally associated with the asymptotic infinity. This 3D manifold is orientable and can be obtained by removing an open ball from $\mathbb{R}^3$ and identifying antipodal points on the boundary sphere. In Einstein gravity, geons can exist only in the presence of exotic matter violating the null energy condition. For example, a self-interacting phantom scalar field has nowadays become a useful mathematical tool in models of astrophysics and cosmology. In particular, some models of dark energy and dark matter are based on phantom scalar fields with special self-interaction potentials.

In what follows, we will restrict our consideration only to static spherically symmetric geons that are supported by a phantom scalar field and do not have an event horizon. In this case, there exists a more convenient presentation of a geon as the quotient by the group $\mathbb{Z}_2$ of a spherically symmetric traversable wormhole~\cite{BronnikovFabris2006, BronnikovMelnikov2007, BronnikovSushkov2011, Bolokhov2012, BronnikovKorolyov2017} that possesses the reflectional symmetry with respect to its throat. The nonidentical $\mathbb{Z}_2$\;\!-\;\!transformation in spherical coordinates has the form $(t,r,\vartheta,\varphi) \rightarrow(t,-r,\pi-\vartheta,\varphi+\pi)$, $r\in\mathbb{R}$. Formally, such a wormhole and the corresponding geon are interchangeable (and equal for $r>0$) solutions of the Einstein-Klein-Gordon equations but have the different topologies and physical interpretations. Wormholes are assumed to have emerged as global primordial structures that connect two large regions of the Universe, while geons are particlelike objects which could be created in the early Universe~\cite{Dowker1998} and now could be of various size from the Planck length up to the galaxy scale. For this reason, it seems preferable to regard geons as potentially observable astrophysical objects~\cite{Kardashev2009}.

There is a natural question whether supermassive black holes and not geons (or wormholes) are placed in the centers of massive galaxies, or in more general context, how we can distinguish geons from black holes~\cite{Li2014}. In this paper, we study the distinctive features of the motion of massive test particles in geon spacetimes of various types. In Section~\ref{Sec2} we introduce the appropriate mathematical technique for describing geons and give some concrete examples. Section~\ref{Sec3} is devoted to studying the bounded orbits in the spirit of the Kepler problem.

\section{Basic equations and types of topological geons}\label{Sec2}

Mathematically, self-gravitating phantom scalar fields are described by the action (in geometric units, $G=c=1$)
\begin{equation}\label{action}
\Sigma=\frac{1}{8\pi}\int\! \left(-\frac{1}{2}S - \langle d\phi,d\phi\rangle-2V(\phi)\right) \sqrt[]{|g|}\,d^{\,4}x\,,
\end{equation}
where $S$ is the scalar curvature, $\phi$ the scalar field, $V(\phi)$ the self-interaction potential, and the sign of the scalar field kinetic term is negative.
It is convenient, for our purposes, to write the metric of a spherically symmetric spacetime in the form
\begin{equation}\label{metric}
ds^2=A dt^2-\frac{\,dr^2}{\,A}-C^{2} \left(d\theta^2+\sin^{2}\theta\,d\varphi^2\right)\,,
\end{equation}
where $A$, $C$ and $\phi$ are functions of the radial coordinate $r$ with the range from $-\infty$ to $\infty$. Then, in the orthonormal basis associated with the metric~(\ref{metric}), the independent Einstein-Klein-Gordon equations for the action~(\ref{action}) are
\begin{eqnarray}\label{equation00}
2A\,\frac{\,C''}{C} + A\,\frac{{\,C'}^2}{C^2} + A'\,\frac{\,C'}{C} - \frac{1}{\,C^2} &=& A{\phi'}^2 -2V\,, \\
\label{equation11}
A'\,\frac{\,C'}{C} + A\frac{{\,C'}^2}{C^2} - \frac{1}{\,C^2} &=& - A{\phi'}^2 -2V\,, \\ \label{equationKG}
A\phi'' + \phi'\!\left(\!A'+2A\,\frac{\,C'}{C}\right) + \frac{dV}{d\phi} &=& 0\,,
\end{eqnarray}
where a prime denotes differentiation with respect to $r$.

From a physical point of view, a scalar field can be considered both as a fundamental natural field and as a phenomenological construction. In any case, we do not know the explicit form of the self-interaction potential. Therefore we assume that the scalar field is nonlinear with an arbitrary potential and have to study geons and their properties using the so-called 'inverse problem method for self-gravitating spherically symmetric scalar fields' developed in Refs.~\cite{BechmannLechtenfeld1995, BronnikovShikin2002}. This method allows one to
construct a 'general solution' of the Einstein-Klein-Gordon equations; it can be represented as quadrature formulae~\cite{Tchemarina2009, Azreg-Ainou2010,Solovyev2012}. Here we use the quadrature formulae in the form~\cite{Solovyev2012}
\vspace{-1ex}
\begin{equation}\label{phi-A}
\phi'= \sqrt{C''/C\,}\,, \quad
A=2C^{2}\!\!\int\limits_{\!r}^{\,\,\infty}\! \frac{\,r}{C^4}\,dr\,,\quad
\widetilde{V}(r)=\frac{1}{2C^2}\!
\left(\!1-3{C'}^2A- CC''A + 2r\,\frac{\,C'}{C}\right).
\end{equation}
It can be directly verified that equations~(\ref{equation00})--(\ref{equationKG}) hold identically for $\phi'$, $A$, and $\widetilde{V}$ defined by~(\ref{phi-A}).

These quadrature formulae give us rich possibilities to model scalar hairy configurations in terms of the area metric function $C(r)$ with the asymptotic expansion $C(r)=r+3m+o(1)$. Choosing $C(r)$ to be even and positive, and such that $C''>0$ for all $r$, one can find $\phi(r)$, $A(r)$, and $\widetilde{V}(r)$; then, if $\phi(r)$ is monotonic, one finds the potential $V(\phi)$; however, each solution to the system~(\ref{equation00})--(\ref{equationKG}) satisfies~(\ref{phi-A}) regardless of the monotonicity of $\phi(r)$. The requirement for $C(r)$ to be even eliminates the ambiguity in the choice of the radial coordinate so that $m$ is the Schwarzschild mass. The formulae~(\ref{equation00})--(\ref{phi-A}) are invariant under the scale transformation $r\rightarrow{}r/\alpha$, $C\rightarrow{}C/\alpha$, $V\rightarrow{}\alpha{}V$, so that every time we can choose an arbitrary scale (e.g., take $m=1$).

\begin{figure}[ht!]
  \centering
  \begin{minipage}{0.43\textwidth}
    \includegraphics[width=\textwidth]{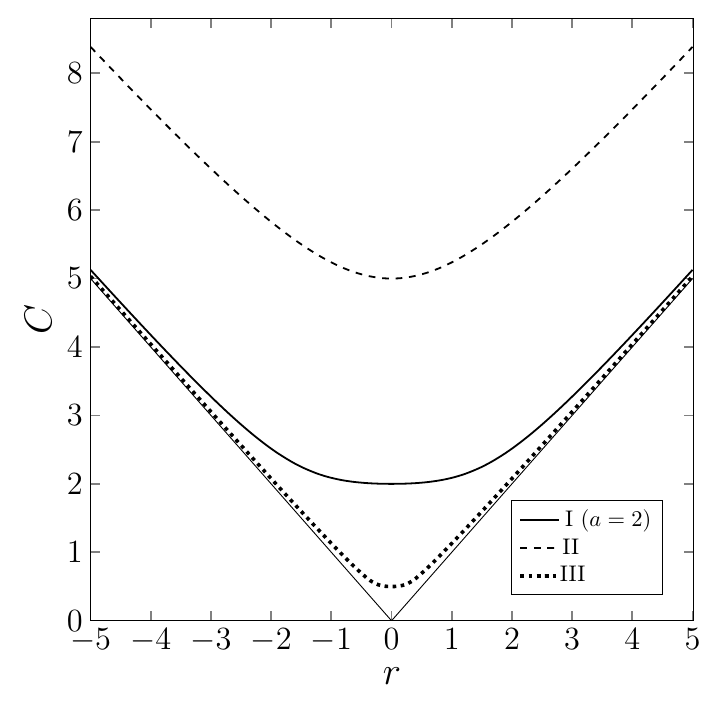}\\
  \end{minipage}\quad\quad\quad
  %%%%%%%%%%%%%%%%%%%%%%%
\begin{minipage}{0.41\textwidth}
    \includegraphics[width=\textwidth]{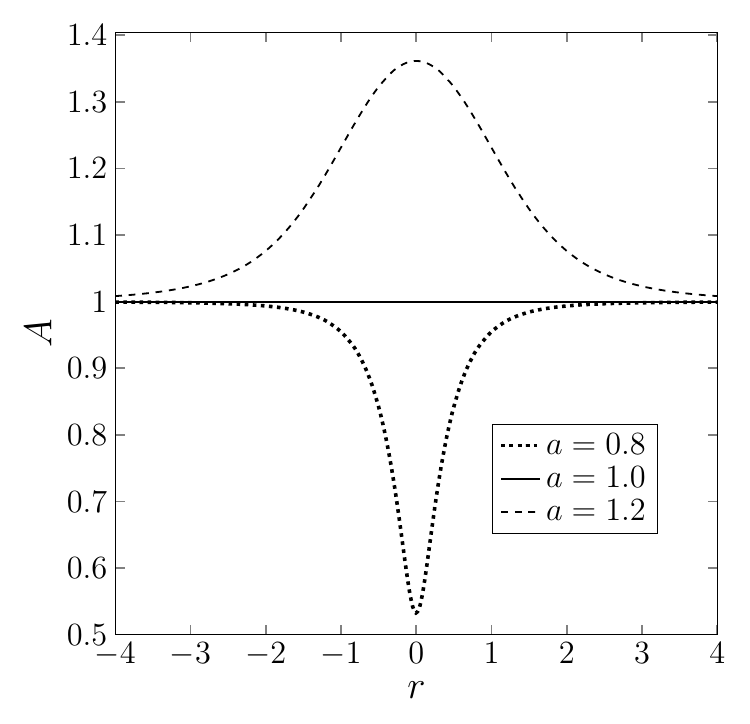}\\
  \end{minipage}
 \begin{minipage}{0.42\textwidth}
    \includegraphics[width=\textwidth]{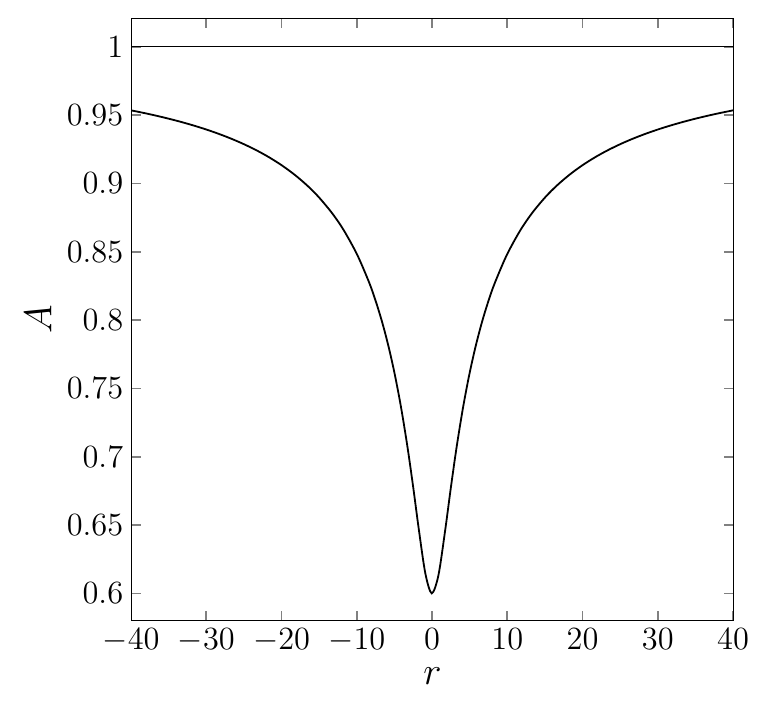}\\
  \end{minipage}\quad\quad\quad
  \begin{minipage}{0.42\textwidth}
    \includegraphics[width=\textwidth]{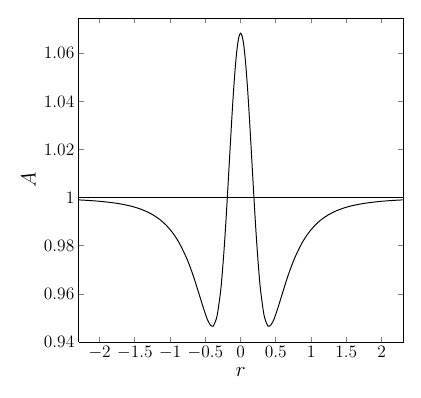}\\
  \end{minipage}
  \caption{Top-left plot: the function $C$ for the cases I, II, and III in~(\ref{C}). Top-right plot: $A(r)$ for the case I; the value $a=1.0$ corresponds to the Ellis solution~\cite{Ellis}. Bottom plots: $A(r)$ for the cases II (left) and III (right)}
  \label{F1}
\end{figure}

There exist, as opposed to black holes, a variety of geons with qualitatively different behaviors of the metric function $A(r)$ near the geon surface (or the wormhole throat). In order to discuss these principal differences, we consider the three illustrative examples labeled by I, II, and III:
\begin{equation}\label{C}
(\mathrm{I})\; C\!=\!(r^4+2r^2+a^4)^{1/4},\quad (\mathrm{II})\; C\!=\!(r^2+4)^{1/2}+3\,,\quad (\mathrm{III})\; C\!=\!\left(\!r^6+r^4+\frac{r^2}{16}+\frac{1}{64}\! \right)^{\!\!1/6}\!\!.
\end{equation}
The geons I and III have a zero Schwarzschild mass, while for the geon II $m=1$. For the cases I and II, the function $C(r)$ has the same form, $C=\sqrt{r^2+a^2\,}+3m$, so from~(\ref{phi-A}) we find $A=1-2m/C$. Note that for any admissible choice of $C(r)$, we have $A=1-2m/r+o(1/r)$ as $r\rightarrow\infty$ and the potential is positive on the geon surface ($r=0$) and negative at spatial infinity, $V\sim-3m/r^3$ as $r\rightarrow\infty$. For the cases I, II, and III, the functions $C(r)$ and $A(r)$ are plotted in Figure~\ref{F1}.

We see that there exist at least three types of geons distinguish by the form of $A$ near the geon surface. The metric function $A$ of the first or the second type has, respectively, only one minimum or maximum that placed on the geon surface (or the wormhole throat); $A$ of the third type has a maximum at $r=0$ and at least one minimum at some point $r=r*>0$.

\begin{figure}[!h]
  \centering
  \begin{minipage}{0.48\textwidth}
    \includegraphics[width=\textwidth]{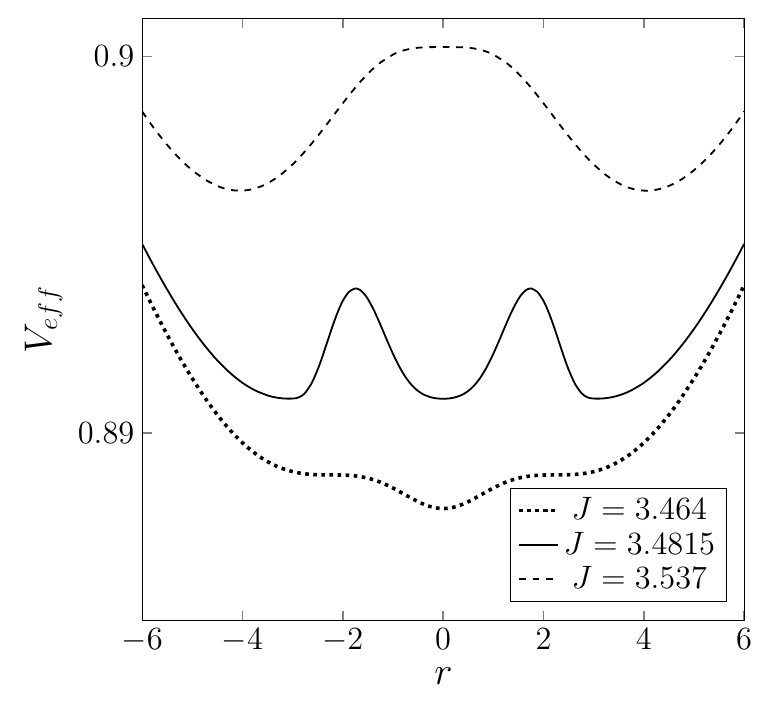}\\
  \end{minipage}
  %%%%%%%%%%%%%%%%%%%%%%%
 \begin{minipage}{0.48\textwidth}
    \includegraphics[width=\textwidth]{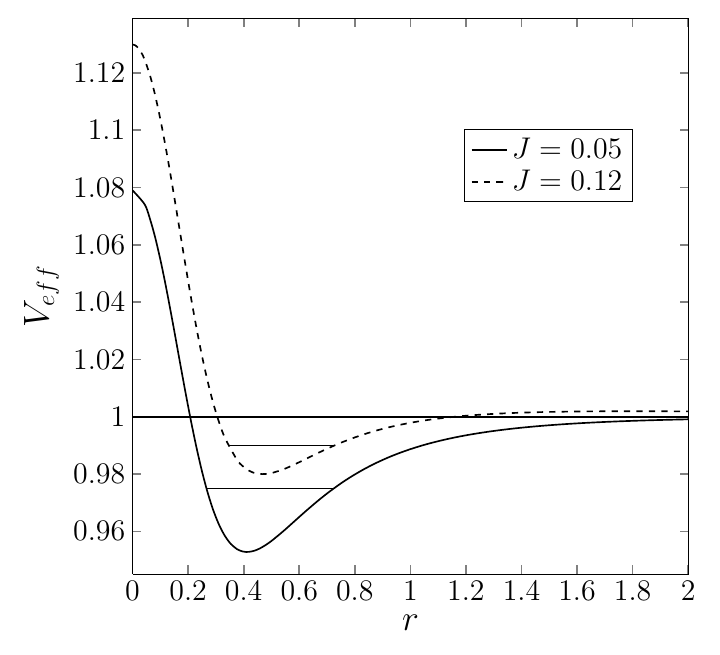}\\
  \end{minipage}
  %%%%%%%%%%%%%%%%%%%%%%%
  \begin{minipage}{0.48\textwidth}
    \includegraphics[width=\textwidth]{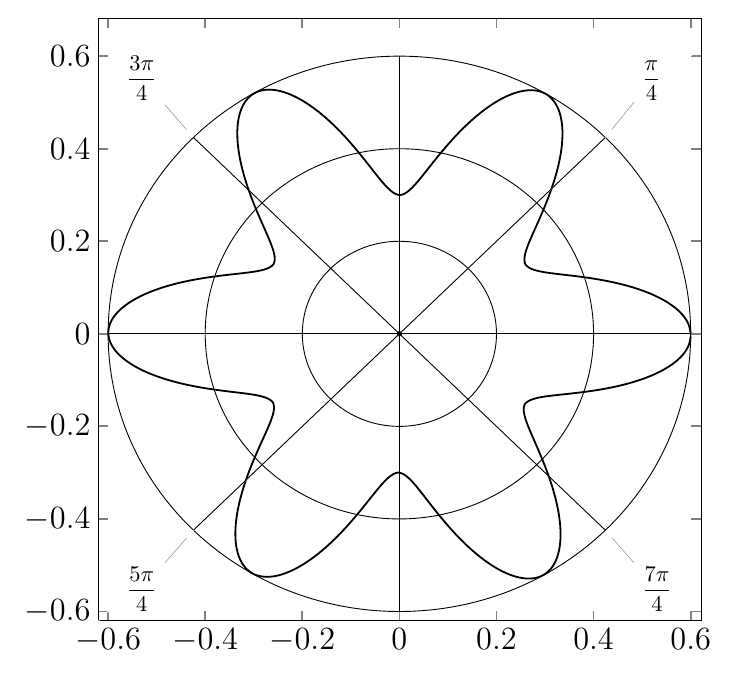}\\
  \end{minipage}
  %%%%%%%%%%%%%%%%%%%%%%%
\begin{minipage}{0.48\textwidth}
    \includegraphics[width=\textwidth]{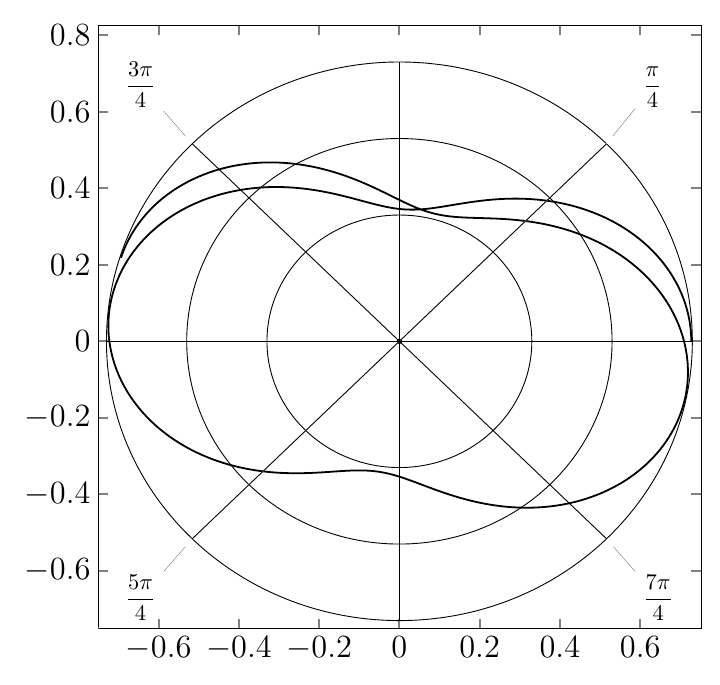}\\
  \end{minipage}
  \caption{Top-left plot: the effective potential for the case II. Top-right plot: the effective potential for the case III. Bottom plots: two trajectories with $E^2=0.965,\;J=0.05$ (left) and $E^2=0.99,\;J=0.12$ (right) for the case III.}
  \label{F2}
\end{figure}

\section{Bounded orbits around topological geons}
\label{Sec3}

The objects of our primary interest are minimums of the metric function $A$ because the first integrals and the effective potential for a massive test particle are~\cite{Nikonov2016}
\begin{equation}\label{Veff}
\frac{dt}{ds}=\frac{E}{A}\,,\quad
\frac{d\varphi}{ds}=\frac{J}{\,C^2}\,,\quad
\left(\frac{dr}{ds}\right)^{\!2}=  E^2- {V\vphantom{\underline{A}}}_{\!\!eff},\quad
{V\vphantom{\underline{A}}}_{\!\!eff}= A\left(1+\frac{\,J^2}{\,C^2}\right),
\end{equation}
where $E$ and $J$ denote, respectively, the 'specific energy' and the 'specific angular momentum' of the particle. A minimum of $A(r)$ guarantees that there exist bounded orbits and stable circular orbits in its neighborhood, including the stable orbits with $J=0$ for particles at rest. Figure~\ref{F2} shows, first, that the effective potential of a geon with $m>0$ can have a minimum on the geon surface and at least one minimum at some point $r=r*>0$ (in the bounded interval, $3.464<J<3.537$). Second, there exists orbits that are impossible in neighborhoods of black holes. The angular velocities of test particles in the circular orbits for the cases II and III, as measured by a static observer at infinity, are given in the table below. If a test particle is in a potential well with nonzero radius $r=r^*>0$ and $E^2={V\vphantom{\underline{A}}}_{\!\!eff}(r^*)$, then the particle has a zero angular velocity (formally, in the stable circular 'orbit'). The similar potential wells were also found in other gravitating systems~\cite{Pugliese2010, Vieira2013}. On the contrary, a test particle can have an arbitrary angular velocity in the stable circular orbit on the geon surface.

\begin{center}
\begin{table}[!h]
\label{T}
\caption{The angular velocity $\omega$ of a test particle in the circular orbits.}
\vspace{1ex}
\centering
\begin{tabular}{|c|c|c|c|c|c|c|}
\hline
 &\multicolumn{3}{|c|}{The case II}&
\multicolumn{3}{c|}{The case III}\\
\hline
$\;r\;$ & 0 & 0  &\quad 3.072\quad\quad &\quad 4.052\quad\quad &\quad 0.412\quad\quad&\quad 0.468\quad\quad \\
\hline
$\;J\;$&\quad 3.464\quad\quad&\quad 3.4815\quad\quad & 3.4815 &3.537 & 0.05 & 0.12 \\
\hline
$\;\omega\;$&0.089&0.089&0.058& 0.049 & 0.124 & 0.259\\
\hline
\end{tabular}
\end{table}
\end{center}

\section{Conclusions}

We show that in neighborhoods of topological geons, test particles can accumulate in spherical layers around potential wells (minimums of the time-time metric function). Near the geon surface, the particles can have arbitrary velocities and the corresponding gas can be at very high temperature, while non-central potential wells attractive particles with zero or small velocities, thus forming cold spherical layers around the geon.
This is a crucial difference from the case of black holes. In the latter case, accretion disks with the innermost stable circular orbit must be observed. We can expect that a massive or supermassive geon can be revealed and identified by the orbital motion of test particles (planets or stars) and the specific behavior of clouds of gas in its neighborhood.


\section*{References }
\begin{thebibliography}{12}

\bibitem{Sorkin1986}
Sorkin R D 1986 \textit{Proc. NATO Advanced Study Institute on Topological Properties and Global Structure of Space-Time, Erice 1985} Ed Bergmann P G and De Sabbata V (New York: Plenum) p 249

\bibitem{BronnikovFabris2006}
Bronnikov K A and Fabris J C 2006
\textit{Phys. Rev. Lett.} \textbf{96}, 251101 (\textit{arXiv:} gr-qc/0511109)

\bibitem{BronnikovMelnikov2007}
Bronnikov K A, Dehnen H and Melnikov V N 2007
\textit{Gen. Rel. Grav.} \textbf{39} 973 (\textit{arXiv:} gr-qc/0611022)

\bibitem{BronnikovSushkov2011}
Bronnikov K A and Sushkov 2010
\textit{Class. Quantum Grav.} \textbf{27}, 095022 (\textit{arXiv:} 1001.3511)

\bibitem{Bolokhov2012}
Bolokhov S V, Bronnikov K A and Skvortsova M V 2012 \textit{Class. Quantum Grav.} \textbf{29} 245006

\bibitem{BronnikovKorolyov2017}
Bronnikov K A and Korolyov P A 2017 \textit{Grav. Cosmol.} \textbf{23} 273 (\textit{arXiv:} 1705.05906)

\bibitem{Dowker1998}
Dowker H F and Sorkin R D 1998 \textit{Class. Quantum Grav.} \textbf{15} 1153 (\textit{arXiv}: gr-qc/9609064)

\bibitem{Kardashev2009}
Kardashev N S 2009 \textit{Phys. Usp.} \textbf{52} 1127

\bibitem{Li2014}
Li Z, Bambi C 2014 \textit{Phys. Rev. D} \textbf{90}, 024071

\bibitem{BechmannLechtenfeld1995}
Bechmann O and Lechtenfeld O 1995
\textit{Class. Quantum Grav.} \textbf{12} 1473 (\textit{arXiv:} gr-qc/9502011)

\bibitem{BronnikovShikin2002}
Bronnikov K A and Shikin G N 2002 \textit{Grav. Cosmol.} \textbf{8} 107 (\textit{arXiv:} gr-qc/0109027)

\bibitem{Tchemarina2009}
Tchemarina Ju V and Tsirulev A N 2009
\textit{Grav. Cosmol.} \textbf{15} 94

\bibitem{Azreg-Ainou2010}
Azreg-A\"inou M 2010 \textit{Gen. Rel. Grav.} \textbf{42} 1427 (\textit{arXiv:} gr-qc/0912.1722)

\bibitem{Solovyev2012}
Solovyev D A and Tsirulev A N 2012
\textit{Class. Quantum Grav.} \textbf{29} 055013

\bibitem{Nikonov2016}
Nikonov V V, Potashov I M and Tsirulev A N 2016 \textit{Math. Model. Geom.} \textbf{5} No 2 http://mmg.tversu.ru

\bibitem{Ellis} Ellis H G 1973
\textit{J. Math. Phys.} \textbf{14} 104 (Errata: 1974 \textit{J. Math. Phys.} \textbf{15} 520)

\bibitem{Pugliese2010}
Pugliese D, Quevedo H, and Ruffini R 2011
\textit{Phys. Rev. D} \textbf{83}, 024021 (\textit{arXiv:} 1003.2687)

\bibitem{Vieira2013}
Vieira R S S, Schee J, Klu\'{z}niak W, Stuchl\'{\i}k Z and Abramowicz M 2014 \textit{Phys. Rev. D} \textbf{90} 024035

\end{thebibliography}
\end{document}